\documentstyle[prl,aps,graphics,floats]{revtex}

\begin{document}

\draft

\title{Double resonance frequency shift in 
a hydrogen maser}

\author{M.A. Humphrey, D.F. Phillips and R.L. Walsworth}
\address{Harvard-Smithsonian Center for Astrophysics, Cambridge, MA 02138}

\date{\today}

\maketitle

\begin{abstract}

We use the dressed atom formalism to calculate the frequency shift in a hydrogen 
maser induced by 
applied radiation near the Zeeman frequency, and find excellent 
agreement with a previous calculation made in the bare atom basis.  
The maser oscillates on the $\Delta F = 1$,  $\Delta
m_F = 0$ hyperfine transition, while the applied field is swept through 
the $F$ = 1, $\Delta m_F = \pm 1$ Zeeman
resonance.  We determine the effect of the applied field on the Zeeman levels 
using the dressed atom picture, and then calculate the maser frequency shift by 
coupling the dressed states to the microwave cavity.  
Qualitatively, the dressed-atom analysis gives a new and
simpler physical interpretation of this double resonance process, which has 
applications in precision hydrogen Zeeman spectroscopy, e.g., in fundamental 
symmetry tests. 

\end{abstract}



\section{Introduction}
\label{sec.intro}

Since its development nearly 40 years ago, the hydrogen maser 
has served as a robust tool capable of
making high precision measurements by utilizing its excellent frequency 
stability \cite{Hmas1,Hmas2,VanAud}.  Hydrogen masers
are currently the most stable active oscillators over intervals of seconds 
to days, with applications including very long baseline interferometry, 
deep space tracking and navigation, and metrology.  
In addition, hydrogen masers have been used to make precision atomic physics 
measurements 
\cite{Hmas3,Hmas4,Hmas5,ronse,UBC} and for sensitive tests of general
relativity \cite{vessot.grav} and quantum mechanics 
\cite{nlquant1,nlquant2}.

\begin{figure}
\begin{center}
\includegraphics{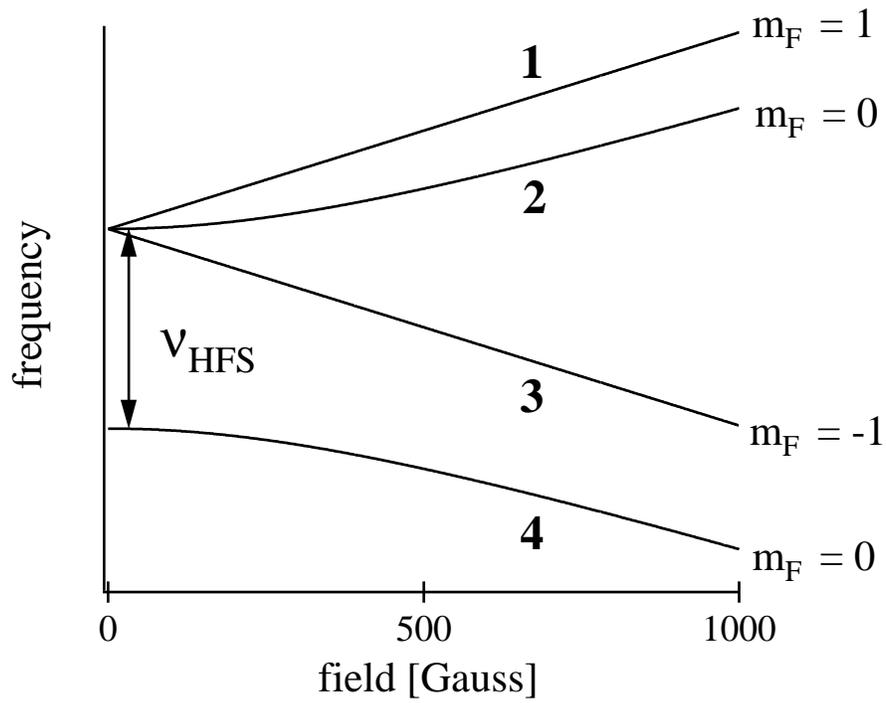}
\caption{Hydrogen hyperfine structure.  A hydrogen maser 
oscillates on the first-order magnetic field-independent $|2\rangle 
\leftrightarrow |4\rangle$ 
hyperfine transition near 1420 MHz.  The maser typically
operates with a static field less than 1 mG.  For these low field 
strengths, the two $F = 1$, $\Delta m_{F}$ = 1 Zeeman frequencies are 
nearly degenerate, and  $\nu_{12} \approx \nu_{23} \approx$ 1 kHz.}
\label{fig.levels}
\end{center}
\end{figure}

\begin{figure}
\begin{center}
\includegraphics{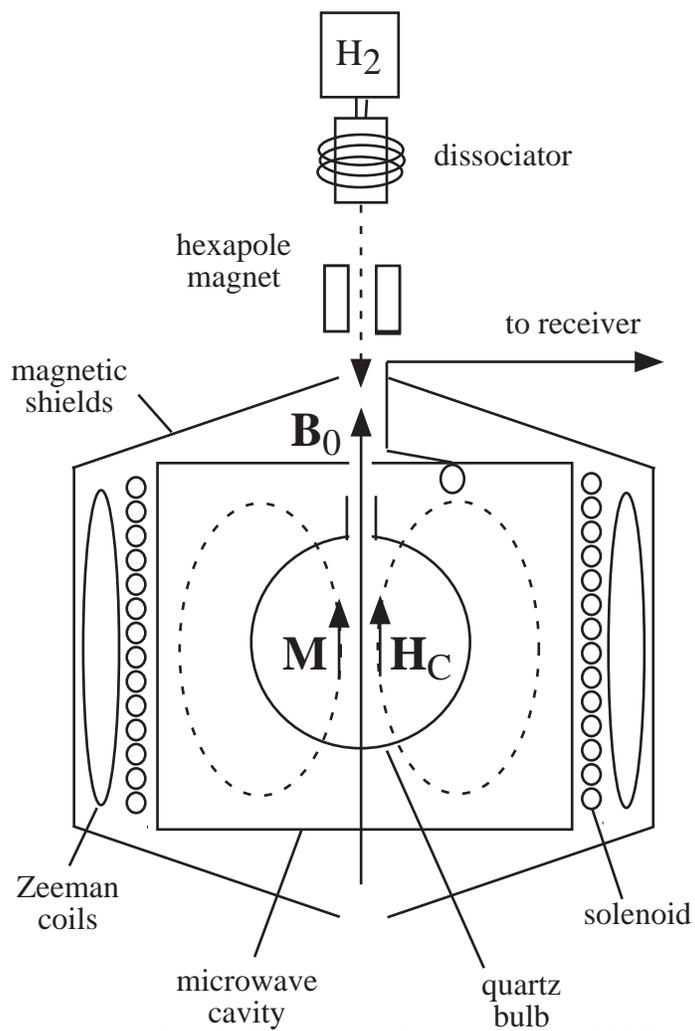}
\caption{Hydrogen maser schematic.  The solenoid generates a weak 
static magnetic field ${\mathbf B}_{0}$ which defines a quantization axis inside 
the 
maser bulb.  The microwave cavity field ${\mathbf H}_{C}$ (dashed 
field lines) and the coherent magnetization ${\mathbf M}$ of the atomic ensemble 
form the coupled actively oscillating system.}
\label{fig.schematic}
\end{center}
\end{figure}

A hydrogen maser operates on the first-order magnetic 
field-independent $\Delta F = 1$, $\Delta m_F = 0$ 
hyperfine transition, between states $|2\rangle$ and $|4\rangle$ of 
the electronic ground state (see Fig. \ref{fig.levels}).  As shown in 
Fig. \ref{fig.schematic}, atoms in states $|1\rangle$ and $|2\rangle$ 
are injected into a storage bulb residing in a microwave cavity tuned 
near the hyperfine frequency, creating the population inversion 
necessary for active maser oscillation.  The microwave field 
stimulates a small, coherent magnetization in the atomic ensemble, and this 
magnetization acts as a source to increase the microwave field.  
With sufficiently high atomic flux and low cavity losses, 
this feedback induces active maser oscillation.  The maser 
frequency is very stable; a well engineered hydrogen maser can have 
fractional frequency stability on the order of 1$\times$10$^{-15}$ over intervals 
of a few hours.   

Hydrogen masers can also be used as sensitive probes of the $F = 1$, 
$\Delta m_F = \pm 1$ Zeeman resonances through a double resonance technique 
\cite{andresen}, 
in which an oscillating magnetic field tuned near the atomic Zeeman resonance 
shifts the $\Delta F = 1$, $\Delta m_F = 0$ maser frequency.  
At low static magnetic fields, this maser frequency shift is an antisymmetric 
function of the detuning of the applied oscillating field from the 
Zeeman resonance.  Thus, by observing the antisymmetric pulling of the 
otherwise stable maser frequency, the hydrogen Zeeman frequency can be determined 
with high precision. 

An early investigation of atomic double resonance was made by 
Ramsey \cite{ramsey.dr}, who calculated the frequency 
shift between two levels coupled by radiation to other levels.  This calculation 
treated the problem perturbatively to first order in the coupling field strength, 
and it neglected damping.  Later, Andresen \cite{andresen,long.andresen} 
calculated the frequency shift in a hydrogen maser due to an applied
field oscillating near the $F = 1$ Zeeman frequency (see Sec. 
\ref{sec.frequency}), and then measured this double resonance effect 
using a hydrogen maser, finding reasonable agreement with the calculation.  
Andresen's calculation employed a bare atom basis, treating the
problem to second order in the applied field strength and including 
phenomenological damping terms, but used an oversimplified description of 
spin-exchange relaxation.  Savard \cite{savard} revisited the problem with a 
more realistic spin-exchange relaxation description
\cite{bender.se}, and found a small correction to the earlier work.

Although the work of Andresen and Savard provides a complete 
description for the double resonance maser frequency shift,
intuitive understanding is obscured by the length of the 
calculations and the use of the bare atom basis.
In particular, these works demonstrate that the amplitude 
of the antisymmetric maser frequency shift is directly proportional to the 
electronic polarization of the masing atomic ensemble.  
The maser frequency shift vanishes as this polarization goes to zero.  
The previous bare atom analyses provide no physical interpretation of 
ths effect.

Since the dressed atom formalism \cite{api} often adds physical 
insight to the understanding of the interaction of matter and 
radiation, we apply it here to the double resonance frequency shift in 
a hydrogen maser.  In a two step process, we first use the dressed atom 
picture to determine the effect of the applied Zeeman radiation on the atomic 
states.  Then, we analyze the effect of the microwave cavity field on
the dressed states and determine the maser frequency shift.  We find 
excellent quantitative agreement between the dressed atom calculation and
the previous bare atom result.  We exploit the dressed atom 
calculation to offer a simple physical interpretation of the double resonance 
hydrogen maser frequency  shift, including a straightforward explanation of 
the polarization dependence.  We conclude by noting the experimental application 
of the double resonance effect to perform improved, high precision hydrogen 
Zeeman spectroscopy.

\section{Maser oscillation frequency}
\label{sec.frequency}  

In a hydrogen maser (Fig. \ref{fig.schematic}), molecular hydrogen is dissociated in an rf
discharge and a thermal beam of hydrogen atoms is formed.  
A hexapole state selecting magnet focuses the higher energy, 
low-field-seeking hyperfine states ($|1\rangle$ and $|2\rangle$) into 
a quartz maser bulb at about $10^{12}$ atoms/sec.  
Inside the bulb (volume $\approx 10^{3}$ cm$^{3}$), the atoms travel 
ballistically for about 1 second before escaping, making $\approx 10^{4}$ 
collisions with the bulb wall.  A Teflon coating reduces the atom-wall 
interaction and thus inhibits decoherence of the 
masing atomic ensemble by wall collisions.  The maser bulb is 
centered inside a cylindrical TE$_{011}$ microwave cavity 
resonant with the 1420 MHz hyperfine transition.  
As described above, the microwave field and 
atomic ensemble form a coupled system with positive feedback 
that will actively oscillate near the $\Delta F = 1$, $\Delta m_F = 0$ hyperfine 
frequency if there is a sufficiently high atomic flux into the maser bulb.  
Since the atoms are confined to a region of uniform microwave field 
phase, their velocity is effectively averaged to zero over the interaction time 
with the microwave field, and first-order Doppler effects are eliminated.  
The maser signal is inductively coupled out of the microwave cavity 
and amplified with an external receiver.  Surrounding the cavity, a solenoid 
produces a weak static magnetic field to maintain the quantization axis inside 
the maser bulb, and a pair of Helmholtz coils produces the oscillating transverse 
magnetic field that drives the $F=1$ Zeeman transitions.  
The cavity, solenoid, and Zeeman drive coils
are all enclosed within several layers of magnetic shielding.

To calculate the maser oscillation frequency, we first determine the relationship between 
the microwave field and a source magnetization.  
If we assume that a single cavity mode is dominant within the microwave cavity,
we may write the microwave magnetic field as
$\mathbf {H}(\mathbf {r},\omega) = \sqrt{4\pi} p_{C}(\omega) 
\mathbf {H}_{C}(\mathbf {r})$, where $p_{C}$
characterizes the frequency- (i.e.\ time-) dependent amplitude and 
$\mathbf{H}_{C}$ represents 
the time-independent spatial variation of the mode.  
A straightforward application of Maxwell's equations relates 
$p_{C}(\omega)$ to a source magnetization M($\omega$): 
  \begin{equation}
  \label{eqn.cav}
  (\omega_{C}^{2} + \frac{i\omega_{C} \omega }{Q_{C}} - \omega^2) p_{C}(\omega) 
  = \sqrt{4\pi} \omega^2 \langle
  H_{C}^{(z)} \rangle_{b} V_{b} M(\omega)
  \end{equation}  
where $\omega$ is the maser frequency, $\omega_C$ is the cavity frequency, 
$\langle H_{C}^{(z)} \rangle_{b}$ is the average of the microwave field's 
z-component over the bulb, and $V_b$ is
the volume of the maser bulb.  The second term on the left has been 
introduced phenomenologically to account for cavity losses 
characterized by $Q_{C}$, the cavity quality factor.  The source magnetization is 
produced by 
the atomic ensemble, and is given by the expectation value of the 
magnetic dipole operator (neglecting the term precessing at $-\omega$),
  \begin{equation}
 	\label{eqn.mag}
  M(\omega)=N \ \langle \hat{\mu} \rangle = N \ Tr(\hat{\rho} 
  \hat{\mu})=N\mu_{24}\rho_{42}(\omega)
  \end{equation} 
where $N$ is the atomic density and $\mu_{24}$ is the hyperfine transition 
dipole matrix element,
$\mu_{24}\approx -\mu_{B}$.  The atomic coherence $\rho_{42}(\omega)$ is 
found by solving the Bloch equations
  \begin{equation}
  \label{eqn.barebloch}
  \dot{\rho} = \frac{i}{\hbar}[\rho,H_0] + \frac{i}{\hbar}[\rho,H_{int}] + 
  \dot{\rho}_{flux} +
  \dot{\rho}_{relax}
  \end{equation}
in steady state.  The unperturbed Hamiltonian, $H_{0}$, includes only 
the atomic state energies (given by the hyperfine and Zeeman 
interactions), and the interaction Hamiltonian, $H_{int}$, 
describes the effect of the microwave cavity field, which couples 
states $|2\rangle$ and $|4\rangle$.  For perfect 
hexapole state selection, the flux term is written
  \begin{equation}
  \label{eqn.bareflux}
     \dot{\rho}_{flux} = \frac{r}{2} \left( |1 \rangle\langle 1| + 
     |2\rangle\langle 2| \right) - r \rho,
  \end{equation}
where the positive terms account for the injection of atoms in states 
$|1\rangle$ and 
$|2\rangle$ at rate $r$, and the last term accounts for bulb escape.  The 
relaxation term, $\dot{\rho}_{relax}$, phenomenologically describes population 
and coherence relaxation due to 
wall collisions, magnetic field inhomogeneities, and spin-exchange collisions.

In the absence of Zeeman radiation and for small detuning of the 
cavity from the $|2\rangle \leftrightarrow |4\rangle$ hyperfine 
transition frequency, $\omega_{24}$, the maser 
frequency is found using Eqns. \ref{eqn.cav} - \ref{eqn.barebloch} to 
be
  \begin{equation}
  \label{eqn.cavpull}
      \omega = \omega_{24} + \frac{Q_{C}}{Q_{l}} \left( \omega_{C} - 
      \omega_{24} \right)
  \end{equation}
where $Q_{l} = \omega_{24} / \Delta \omega$, the atomic line-Q, 
is on the order of $10^{9}$  \cite{Hmas1}.
Since the cavity-Q is on the order of $10^{4}$, the cavity pulling is highly 
suppressed, 
and the maser oscillates near the atomic hyperfine frequency, $\omega_{24}$.  

\begin{figure}
\begin{center}
\includegraphics{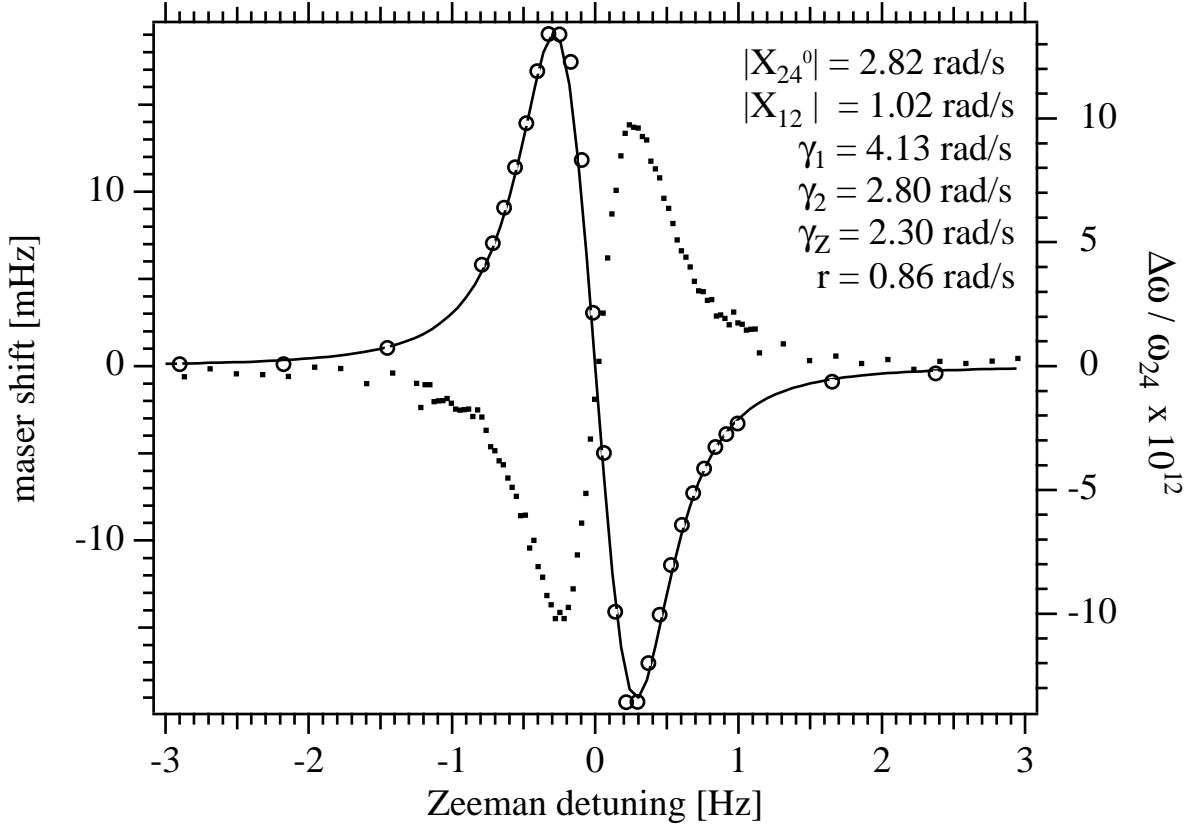}
\caption{Double resonance maser frequency shifts.  The large open circles 
are data taken with an input beam of $|1\rangle$ and $|2\rangle$ hydrogen atoms. 
These are compared with Eqn. \ref{eqn.analytic.andresen} (full curve) using 
the parameter values shown.  The values of $|X_{12}|$ and $\gamma_{Z}$ 
were chosen to fit the data, while the remaining parameters were 
independently measured. 
The experimental error of each measurement is smaller than the circle marking it.
The electronic polarization dependence of the double resonance effect is 
illustrated with the 
dotted data points:  with an input beam of $|2\rangle$ and $|3\rangle$ atoms, 
the shift is inverted.
Note that the maser frequency shifts for the dotted points were scaled up by a 
factor of 10 since 
these data were acquired with a much weaker applied Zeeman field.  
The large variation of the maser frequency shift with 
Zeeman detuning near resonance, along with the excellent maser 
frequency stability, allows the Zeeman frequency ($\approx$ 800 Hz) to be 
determined to 
about 3 mHz in a single scan of the double resonance such as the 
dotted data shown here (requiring $\approx$ 20 minutes of data acquisition).}
\label{fig.andresen}
\end{center}
\end{figure}

In the presence of Zeeman radiation, the maser frequency is shifted, 
as first shown by Andresen \cite{andresen} and measured anew here 
(see Fig. \ref{fig.andresen}).  His calculation
of the double resonance shift included the applied Zeeman field
in the interaction Hamiltonian but otherwise left unchanged 
the above analysis for the maser frequency.  To second order 
in the Rabi frequency of the applied Zeeman field, $|X_{12}|$, and  
in terms of the unperturbed maser Rabi frequency $|X_{24}^{0}|$, 
atom flow rate $r$, population decay rate $\gamma_{1}$, 
hyperfine decoherence rate $\gamma_{2}$, and Zeeman decoherence rate 
$\gamma_{Z}$, the small static field limit of the maser shift is 
given by \cite{factor.two}
    \begin{equation}
    \label{eqn.analytic.andresen}
    \Delta \omega = - |X_{12}|^{2} 
    \frac{\gamma_{Z}}{r} (\gamma_{1} \gamma_{2}+|X_{24}^{0}|^{2})
    \frac{\delta \ (\rho_{11}^{0} - \rho_{33}^{0})}
    {(\gamma_{Z}^{2} - \delta^{2} + |X_{24}^{0}|^{2})^{2} + 
    (2 \delta \gamma_{Z})^{2}}
    \end{equation} 
where $\delta$ is the detuning of the applied field from the atomic Zeeman 
frequency, and $\rho_{11}^{0} - \rho_{33}^{0} = r / (2 \gamma_{1})$ is the steady 
state 
population difference between states $|1\rangle$ and $|3\rangle$ in the absence 
of an 
applied Zeeman field (following Eqn. 8 of ref. \cite{andresen}).   Physically, 
the population difference between states $|1\rangle$ and $|3\rangle$ represents 
the 
electronic polarization of the hydrogen ensemble \cite{long.andresen}:
  \begin{equation}
	P = \frac{ \langle S_{Z} \rangle}{S}  = 2 \ Tr(\hat{\rho} \hat{S_{Z}}) 
	= \rho_{11} - \rho_{33}.
  \end{equation}
Equation \ref{eqn.analytic.andresen} implies that a steady state 
electronic polarization, and hence a population difference between 
states $|1\rangle$ and $|3\rangle$ injected into the maser bulb, is a 
necessary condition 
for the maser to exhibit a double resonance frequency shift.  
Walsworth et. al. demonstrated this polarization dependence experimentally by 
operating 
a hydrogen maser in three configurations:  (i) with the usual input 
flux of atoms in states $|1\rangle$ and $|2\rangle$; (ii) with a pure input 
flux of atoms in state $|2\rangle$, where the maser frequency shift vanishes; 
and (iii) with an input beam of atoms in states $|2\rangle$ and 
$|3\rangle$, where the maser shift is inverted \cite{ronse}.

For typical applied Zeeman field strengths, the 1420 MHz maser frequency 
is shifted tens of mHz (see Fig. \ref{fig.andresen}),  
a fractional shift of $\approx$ 10$^{-11}$.  
However, the shift is easily resolved because of the excellent fractional
maser frequency stability (parts in $10^{15}$).

\section{Dressed atom calculation}
\label{sec.dressed}

We now consider an alternative approach to calculating the double resonance 
maser frequency shift using the dressed atom picture.  
We retain the Maxwell-Bloch formalism of Eqns. \ref{eqn.cav} - 
\ref{eqn.barebloch}; 
however we determine the steady state coherence $\rho_{42}(\omega)$ in a dressed 
atom basis, including the atomic state energies, the applied Zeeman field, and 
the atom/Zeeman field interaction.  
For simplicity, we assume the static magnetic field is sufficiently low 
that the two $F$ = 1, $\Delta m_F = \pm 1$ Zeeman frequencies are nearly 
degenerate, 
$\omega_{12} - \omega_{23} \ll \gamma_{Z}$, as is the case 
for typical hydrogen maser operation.  We use the simplified spin-exchange 
relaxation model \cite{andresen} and neglect the small spin-exchange correction 
to the double resonance maser frequency shift \cite{savard}.

\subsection{Dressed atom basis}
\label{subsec.basis}

By incorporating the applied Zeeman field into the unperturbed Hamiltonian, it 
takes the 
form $H_{0} = H_{a} + H_{f} + V_{af}$.  The atomic states (defining state $|2 
\rangle$  as energy zero) are described by
    $H_{a} = \hbar\omega_{12}|1 \rangle\langle 1| - \hbar\omega_{23}|3 
    \rangle\langle 3| - \hbar\omega_{24}|4
    \rangle\langle 4|$;
the applied Zeeman field (at frequency $\omega_{Z}$) is described by
$H_{f} = \hbar \omega_{Z} a^{\dagger} a$; and the 
interaction between them is given by
  \begin{equation}
      V_{af} = \hbar g \left( a + a^{\dagger} \right) 
      \left[ |1 \rangle\langle 2| + |2 \rangle\langle 3| + h.c. \right].
  \end{equation}
Here, the Zeeman field creation and annihilation operators are $a^{\dagger}$ and 
$a$, $h.c.$ denotes Hermitian 
conjugate, and $g$ is the single-photon Rabi frequency for the Zeeman 
transitions.  We will use eigenkets with two indices to account for the atomic 
state and 
the number of photons in the Zeeman field, denoted by $n$.  We select four 
of these as our bare atom/Zeeman field basis, 
$\left\{ |1,n-1 \rangle, |2,n \rangle, |3,n+1 \rangle, |4,n\rangle \right\}$,
where the first entry indicates the atomic state and the second entry 
indicates the Zeeman photon number.  We note 
that for a resonant field, $\omega_{Z} = \omega_{12}$, the first three 
basis states are degenerate.  Also, $n \gg 1$ in practice for there to 
be any measureable double resonance maser frequency shift.

In this bare atom/Zeeman field basis, the unperturbed Hamiltonian operator takes 
the following matrix form:
  \begin{equation}
  \label{eqn.matrix}
  H_{0} =  \hbar \left(
    \begin{array}{cccc}
	-\delta & \frac{1}{2} X_{12} & 0 & 0 \\
	\frac{1}{2} X_{12} & 0 & \frac{1}{2} X_{23} & 0 \\
	0 & \frac{1}{2} X_{23} & \delta & 0 \\
	0 & 0 & 0 & -\omega_{24}
    \end{array}
    \right),
  \end{equation}
where $\delta = \omega_{Z} - \omega_{12}$ is the detuning of the 
applied Zeeman field, and
  \begin{equation}
    \frac{X_{12}}{2} = g \sqrt{n} \approx g \sqrt{n+1} = \frac{X_{23}}{2}
 \end{equation}
define the Zeeman field Rabi frequency (the factor of two has been 
inserted to be consistent with our rotating wave approximation
convention; see note \cite{factor.two}).  By diagonalizing $H_{0}$, 
we find new basis states which physically represent the atomic states 
dressed by the applied Zeeman field.  The dressed energy levels 
are the eigenvalues of $H_{0}$: $E_{a} = \hbar 
\Omega$, $E_{b} = 0$, $E_{c} = - \hbar 
\Omega$, and $E_{4} = -\hbar \omega_{24}$, where $\Omega = \sqrt{ \delta^{2} + 
\frac{1}{2}X_{12}^{2}}$
represents the generalized Rabi frequency.  The dressed states are the 
eigenvectors of $H_{0}$:
  \begin{eqnarray}
  \label{eqn.evectors}
       & & |a\rangle = \frac{1}{2} \left( 1 - \frac{\delta}{\Omega} 
        \right) |1,n-1\rangle + \frac{X_{12}}{2 \Omega} |2,n\rangle + 
	\frac{1}{2} \left( 1 + \frac{\delta}{\Omega} 
        \right) |3,n+1\rangle \nonumber \\
       & & |b\rangle = \frac{X_{12}}{2 \Omega} |1,n-1\rangle + 
         \frac{\delta}{\Omega} 
         |2,n\rangle - \frac{X_{12}}{2 \Omega} |3,n+1\rangle \\   
       & & |c\rangle = \frac{1}{2} \left( 1 + \frac{\delta}{\Omega} 
        \right) |1,n-1\rangle - \frac{X_{12}}{2 \Omega} |2,n\rangle + 
	\frac{1}{2} \left( 1 - \frac{\delta}{\Omega} 
        \right) |3,n+1\rangle  \nonumber \\	
       & & |4\rangle = |4,n\rangle. \nonumber 
  \end{eqnarray}
Note that in the limit of large negative $\delta$, $|a\rangle 
\rightarrow |1\rangle$ and $|c\rangle \rightarrow |3\rangle$, while 
in the limit of large positive $\delta$, $|a\rangle 
\rightarrow |3\rangle$ and $|c\rangle \rightarrow |1\rangle$.  This 
will become important in the physical interpretation of the 
maser frequency shift (see Sec. \ref{sec.picture}).
  
An operator $\hat{O}$ transforms between bare atom to dressed atom 
bases as 
$\hat{O^{d}} = T^{-1} \hat{O^{b}} T$, where $T$ is the unitary matrix 
linking the dressed atom and bare atom basis states (coefficients of 
Eqn. \ref{eqn.evectors}).  The dressed and bare atom energies 
and eigenvectors are equivalent for the $F=0$ hyperfine state 
$|4\rangle$ because this state is unaffected by the applied Zeeman 
field.

\subsection{Dressed basis Bloch equations}
\label{subsec.dress-Bloch}

We now couple the dressed states to the microwave cavity using the Bloch 
equations, which remain of the form
  \begin{equation}
  \label{eqn.dressbloch}
      \dot{\rho^{d}} = \frac{i}{\hbar}[\rho^{d},H_0^{d}] + 
      \frac{i}{\hbar}[\rho^{d},H_{int}^{d}] + \dot{\rho^{d}}_{relax} +
  \dot{\rho^{d}}_{flux}.
  \end{equation}
The unperturbed Hamiltonian, $H_{0}$, now accounts for the bare atom energies 
and the applied Zeeman driving  
field, while the microwave cavity field is included in the interaction 
Hamiltonian, $H_{int}$.  Since the dressed states $|a\rangle, |b\rangle$, and 
$|c\rangle$ 
all have a component of the atomic state $|2\rangle$ (see Eqn. 
\ref{eqn.evectors}), the microwave field couples 
state $|4\rangle$ to each:
  \begin{equation}
  \label{eqn.int}
      H_{int}^{d} = \frac{X_{12}}{2 \Omega} H_{24} |a \rangle\langle 4| + 
      \frac{\delta}{\Omega} H_{24} |b \rangle\langle 
      4| - \frac{X_{12}}{2 \Omega} H_{24} |c \rangle\langle 4| + h.c.
  \end{equation}
Note that $H_{24} = \langle 2 | \hat{\mu} \cdot {\bf H_{C}}| 4 
\rangle$ is the only nonzero coupling between bare atom states 
supported by the TE$_{011}$ mode microwave cavity.
To transform the relaxation terms into the dressed basis, 
we make the approximation that all relaxation rates (population 
decay $\gamma_{1}$, hyperfine decoherence $\gamma_{2}$, and Zeeman 
decoherence $\gamma_{Z}$) have the same value, 
$\gamma + r$ ($\gamma$ includes all relaxation exclusive of bulb loss).  
Typically, these rates
are within a factor of two (see the values listed in Fig. \ref{fig.andresen}).  
Then,
  \begin{equation}
    \dot{\rho}_{relax}^{d} = -\gamma \rho^{d} + \frac{\gamma}{4} {\mathbf 1}.
  \end{equation}
In the bare atom basis, the flux term has a very simple form 
(Eqn. \ref{eqn.bareflux}) with no off-diagonal input entries 
since the injected beam has no coherence between the bare atomic states.  
In the dressed basis, however, there is an injected Zeeman coherence, so 
the flux term takes a considerably more complicated form
  \begin{equation}
    \dot{\rho}_{flux}^{d} = \frac{r}{2} F^{d} - r \rho^{d}
  \end{equation}
where $F^{d} = T^{-1} \left( |1 \rangle\langle 1| + |2 \rangle\langle 2| 
\right) T$ has three diagonal and six off-diagonal entries. 

The Bloch equations are most easily handled by moving to the 
interaction picture, given by $\hat{O} = e^{-i \hat{H}_{0}t/\hbar} \
\tilde{\hat{O}} \ e^{i \hat{H}_{0}t/\hbar}$, where $\tilde{\hat{O}}$ is 
an interaction picture operator. 

\subsection{Steady state solution}
\label{subsec.ss}

The 4$\times$4 matrix equation (\ref{eqn.dressbloch}) yields 
sixteen independent equations that we solve in the steady state.  
Then, the populations in the interaction picture are static, 
$\dot{\tilde{\rho}}_{\nu \nu} = 0$, 
and the coherences exhibit sinusoidal precession.  In particular,  
$\tilde{\rho}_{4a} = R_{4a} e^{-i (\Omega - \Delta) t}$, 
$\tilde{\rho}_{4b} = R_{4b} e^{i \Delta t}$, and 
$\tilde{\rho}_{4c} = R_{4c} e^{i (\Omega + \Delta) t}$, 
where the $R_{\mu \nu}$ are time independent, and $\Delta = \omega - 
\omega_{24}$.  The other coherences 
precess at frequencies $\omega_{\mu \nu} = (E_{\mu} - E_{\nu})/\hbar$.
Making these steady state substitutions, the sixteen Bloch differential 
equations transform to a set of time-independent algebraic equations.
We assume that $\omega_{C} = \omega_{24}$, so that the small cavity pulling 
shift vanishes.  The total maser frequency shift is then given by $\Delta$. 

In terms of dressed basis density matrix elements (rotated out of the 
interaction picture), the atomic 
coherence $\rho_{42}(\omega)$ is given by
  \begin{equation} 
  \label{eqn.dress.coh}
	 \rho_{42}(\omega) = 
	 \left( \frac{X_{12}}{2 \Omega} \rho_{4a} + 
             \frac{\delta}{\Omega} \rho_{4b} - 
	     \frac{X_{12}}{2 \Omega} \rho_{4c} \right)
        = \left( \frac{X_{12}}{2 \Omega} R_{4a} + 
             \frac{\delta}{\Omega} R_{4b} - 
	     \frac{X_{12}}{2 \Omega} R_{4c} \right) e^{i \omega t}
  \end{equation}
and the magnetization is found from equation (\ref{eqn.mag}).
Inserting this into equation (\ref{eqn.cav}) we find the following two 
conditions which determine the maser amplitude and oscillation frequency
  \begin{eqnarray}
  \label{eqn.cavcouple}
	 & & Re \left( \frac{X_{12}}{2} R_{4a} + 
          \delta R_{4b} - \frac{X_{12}}{2} R_{4c} \right) = -|X_{24}|
          \left( \frac{2 Q_{C} \Delta }{\omega_{C}} \right) 
	  \left[ \frac{(\gamma + r)^{2}}{r \Omega} \left( \frac {I_{0}}{I_{th}} 
	  \right) \right]^{-1} \\
         & & Im \left( \frac{X_{12}}{2} R_{4a} + 
          \delta R_{4b} - \frac{X_{12}}{2} R_{4c} \right) = -|X_{24}| \left[ 
	  \frac{(\gamma + r)^{2}}{r \Omega} \left( \frac {I_{0}}{I_{th}} 
	  \right) \right]^{-1} \nonumber 
   \end{eqnarray}
where  $I_{0} = r V_{b} N$ is the total atomic flux into the maser 
bulb and $I_{th}$ is the threshold flux for maser oscillation with our 
simplified spin-exchange model \cite{Hmas2}: 
  \begin{equation}
    I_{th}= \frac{\hbar V_{C} (\gamma + r)^{2}}{4 \pi |\mu_{24}|^{2} 
    Q_{C} \eta}.
  \end{equation}
Here $V_{C}$ is the cavity volume and $\eta$ is a dimensionless filling factor 
\cite{Hmas1,Hmas2}.

We numerically solve the time-independent algebraic system of sixteen Bloch 
equations plus equations (\ref{eqn.cavcouple}) to determine 
the maser frequency shift $\Delta$ as a function of Zeeman detuning $\delta$.  
We find excellent agreement with the previous theoretical bare atom analysis 
\cite{andresen}, 
within the approximation of equal population decay and decoherence rates.  
(Note that in practice, these decay rates differ by up to a factor of two).

\section{Physical interpretation}
\label{sec.picture}
 
The dressed state analysis provides a straightforward physical 
interpretation of the double resonance maser frequency shift.  In the 
absence of the applied Zeeman field, atoms injected in bare state $|2\rangle$ 
are the sole source of the magnetization that provides the positive 
feedback needed for active oscillation.  However, when the near-resonant Zeeman 
field is 
applied, it also allows atoms injected in the $m_{F} = \pm$ 1 states 
(bare states $|1\rangle$ and $|3\rangle$) to contribute via a 
two-photon process.  A dressed atom interpretation shows how these 
$m_{F} = \pm$ 1 state atoms can become 
the dominant source of maser magnetization as the applied Zeeman field nears 
resonance.  

\begin{figure}
\begin{center}
\includegraphics{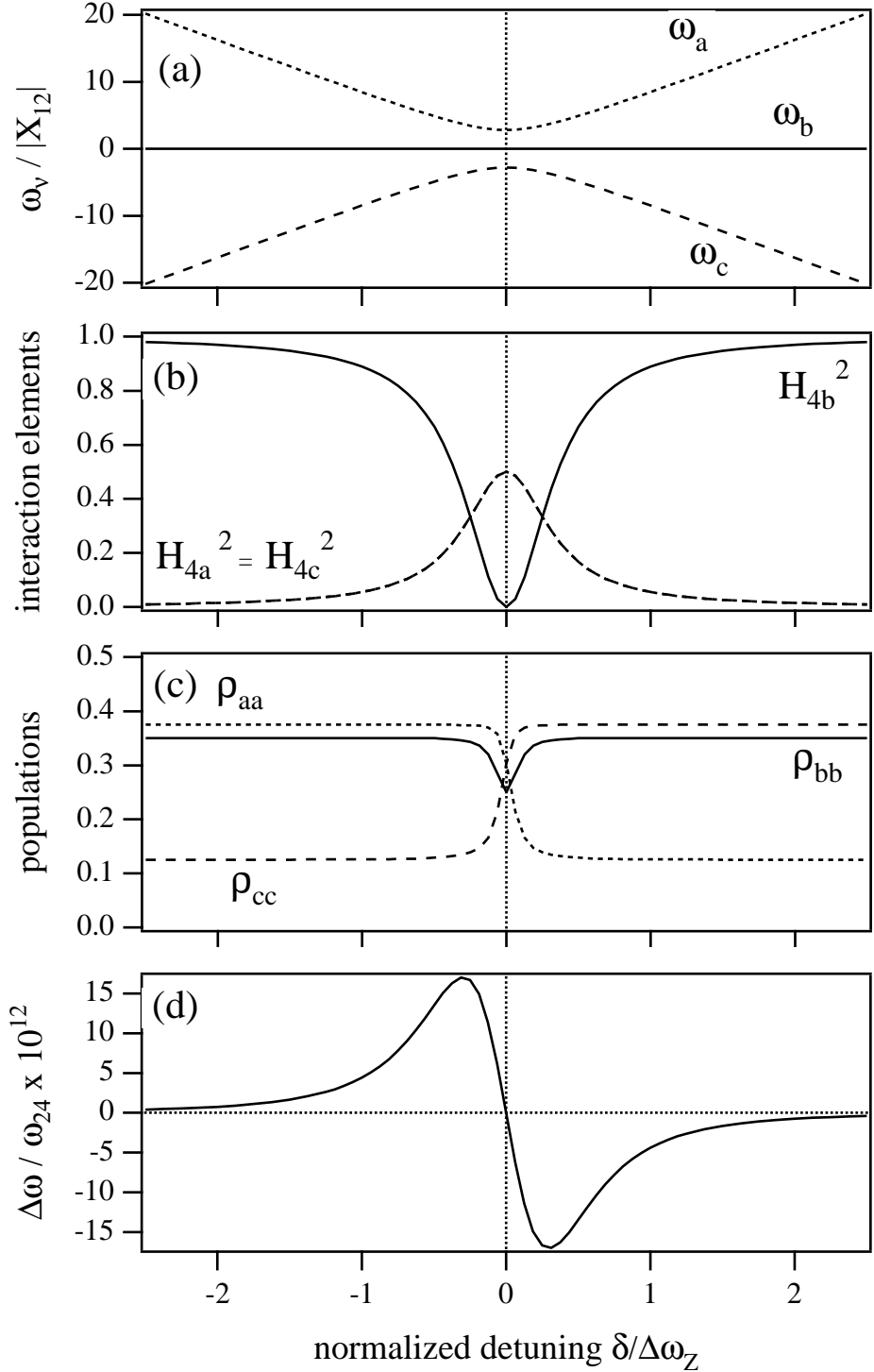}
\caption{Calculated dressed atom quantities plotted against 
detuning of the applied Zeeman field
(in units of Zeeman linewidth, $\Delta \omega_{Z} = 2 \gamma_{Z}$). 
The dotted, full, and dashed curves correspond to dressed states $|a\rangle$, 
$|b\rangle$, and $|c\rangle$, respectively.
(a) Dressed atom frequencies normalized to the Zeeman Rabi frequency.  
(b) Interacton Hamiltonian matrix elements (squared) from equation 
(\ref {eqn.int}) 
in units of $\langle 2 | \hat{\mu} \cdot {\bf H_{C}}| 4 \rangle^{2}$.  
(c) Steady state populations of dressed states.  
(d) Fractional double resonance maser frequency shift.}
\label{fig.picture}
\end{center}
\end{figure}

Viewed from the dressed atom basis, three factors contribute
to this interpretation.  First, as shown in Fig. \ref{fig.picture}(a),
the applied Zeeman field shifts the energies of the two dressed 
levels $|a\rangle$ and $|c\rangle$ symmetrically relative to level 
$|b\rangle$, which remains unperturbed.
Second, near the Zeeman resonance, the $\Delta F = 1$
dipole coupling $H_{4b}^{2} = \langle 4 | \hat{\mu} \cdot {\bf 
H_{C}}| b \rangle^{2}$ vanishes while $H_{4a}^{2}$ and $H_{4c}^{2}$ 
become equally dominant, as shown in Fig. \ref{fig.picture}(b).  
Third, below resonance ($\delta < 0$) the steady state population of state 
$|a\rangle$ is 
greater than that of state $|c\rangle$ ($\rho_{aa} > \rho_{cc}$), while above 
resonance ($\delta > 0$)
the opposite is true ($\rho_{cc} > \rho_{aa}$), as shown in Fig. 
\ref{fig.picture}(c).  
These dressed state population differences arise from the fact that atoms in 
bare state $|1\rangle$ are injected into the maser while those in bare state 
$|3\rangle$ are 
not, under normal operation, so in the steady state $\rho_{11} > \rho_{33}$.  
For large
negative Zeeman detunings, $|a\rangle \rightarrow |1\rangle$ and $|c\rangle 
\rightarrow|3\rangle$ (see discussion following Eqn. \ref{eqn.evectors}).  
The opposite holds for positive detuning, where $|a\rangle \rightarrow 
|3\rangle$ and $|c\rangle \rightarrow|1\rangle$.    

These three ingredients combine to create the double resonance shift 
of the maser frequency, shown in Fig. \ref{fig.picture}(d).  
For small negative Zeeman detunings ($|\delta| < 2 \gamma_{Z}$), the excess of 
$\rho_{aa}$ over $\rho_{cc}$ and the relatively small size of 
$H_{4b}^{2}$ leads to maser oscillation primarily on the 
$|a\rangle \leftrightarrow |4\rangle$ transition.
That is, atoms injected into the maser cavity in the bare state $|1\rangle$ 
contribute significantly to the maser oscillation via a two-photon 
process: one Zeeman transition photon and one microwave photon within 
the resonant cavity linewidth.  This $|a\rangle \leftrightarrow |4\rangle$ 
transition is at a slightly higher frequency than in the unperturbed 
(no applied field) maser, so the maser frequency is increased.  
Conversely, for small positive Zeeman detunings ($\delta < 2 \gamma_{Z}$), 
the maser oscillates preferentially on the  
$|c\rangle \leftrightarrow |4\rangle$ transition, and the maser 
frequency is decreased.  
For larger Zeeman detunings (positive or negative), the coupling of 
state $|4\rangle$ to unshifted dressed state $|b\rangle$ becomes 
dominant, and the magnitude of the frequency shift is reduced.
For zero Zeeman detuning, dressed states $|a\rangle$ and $|c\rangle$ 
are equally populated in the steady state and the maser frequency shift 
exactly vanishes.  

Injection of an electronic polarization into the maser bulb is needed 
for the applied Zeeman field to induce a maser frequency shift.  
Since $\omega_{a}$ and $\omega_{c}$ are spaced equally about the unperturbed 
maser frequency $\omega_{b}$, and since $H_{4a}^{2} = H_{4c}^{2}$, a necessary 
condition for a maser shift is a difference in the steady state values 
of $\rho_{aa}$ and $\rho_{cc}$, which is a direct consequence of a difference in 
the injected populations of bare states $|1\rangle$ and $|3\rangle$, i.e., a net 
electronic polarization.

\section{application}
\label{sec.apps}

The double resonance hydrogen maser technique was originally studied for use in 
autotuning the maser cavity\cite{andresen}.  In 
addition to the double resonance frequency shift, there is a
cavity pulling shift for a mistuned maser cavity, with magnitude 
dependent on the linewidth of the hyperfine transitions, 
through the line-Q (see Eqn. \ref{eqn.cavpull}).  The applied Zeeman radiation 
depletes the population of bare state $|2 \rangle$, thereby increasing the 
linewidth of the 
hyperfine transition.  Andresen \cite{andresen} showed that the cavity can be 
tuned to the atomic frequency by modulating the hyperfine linewidth 
with applied Zeeman radiation and adjusting the cavity frequency such 
that there is no modulation of the maser frequency.  However, this 
method requires accurate setting of the applied Zeeman field to the Zeeman 
resonance (i.e. $\delta$ = 0).

The double resonance technique can also 
be used for precision Zeeman spectroscopy in a hydrogen maser.  
Traditionally, the Zeeman frequency in a hydrogen maser operating at 
low magnetic fields is measured by sweeping an audio frequency field 
through the Zeeman resonance and monitoring the maser power.  The 
power is diminished near resonance with a Lorentzian shape with a 
width on the order of 1 Hz.  Typical resolution of the Zeeman 
frequency with this technique is about 100 mHz.
However, by utilizing the sharp, antisymmetric profile of the 
double resonance frequency shift, we were able to determine 
the hydrogen Zeeman frequency with a resolution of about 1 mHz.
Recently we used this double resonance technique in a search for Lorentz 
symmetry violation of the hydrogen atom's electron and proton spin 
\cite{lli.Hexp}, motivated by a general extension of the standard 
model of elementary particle physics \cite{lli.Hthe}.

\section{conclusion}
\label{sec.con}

We used the dressed atom formalism to calculate the frequency shift 
of a hydrogen maser induced by an applied field near the $F = 1$, $\Delta 
m_{F} = \pm 1$ Zeeman transition frequency.  The result is in 
excellent quantitative agreement with previous bare atom basis 
calculations \cite{andresen,savard}, within a simplified spin-exchange 
approximation 
and with equal decay rates for all populations and coherences.  
The dressed atom picture provides a simple physical understanding of the double 
resonance 
frequency shift, and in particular, the atomic polarization dependence 
of the frequency shift.  The double resonance technique can be 
employed in precision spectroscopy of the hydrogen Zeeman frequency, 
e.g. in a test of Lorentz symmetry of the standard model 
\cite{lli.Hexp,lli.Hthe}.

\section{acknowledgments}

We thank Mikhail Luken, Ed Mattison and Robert Vessot for useful 
discussion.  This work was supported by NASA grant NAGS-1434.  
MAH thanks NASA for support under the Graduate Student 
Researchers Program.


\bibliographystyle{prsty}

\begin{thebibliography}{10}
\bibitem{factor.two}
We have introduced a factor of $\frac{1}{2}$ to the values for the Rabi
  frequencies $|X_{24}|$, $|X_{12}|$, and $|X_{23}|$ to account for the use of
  the rotating wave approximation.
\end{thebibliography}


\end{document}